# Effective roles of the head and flagellum in sperm motility

Robert L. Scott[1], Shobitha Unnikrishnan[1], Ayomide J. Bolaji[1], Luke Achinger[2], Tomer Avidor-Reiss[2], Kenta Ishimoto[3], and Chih-Kuan Tung[1]✉

[1]Department of Physics, North Carolina A&T State University, Greensboro, NC, USA
[2]Department of Molecular, Cellular, and Developmental Biology, The University of Toledo, Toledo, OH, USA
[3]Department of Mathematics, Kyoto University, Kyoto, Japan
✉ email: ctung@ncat.edu

In the low Reynolds number fluid environment that microswimmers encounter, back-and-forth motion cannot lead to net displacement. In mammalian sperm, the mechanical wave propagating along their single flagellum breaks the cancellation between back-and-forth motion and, therefore, is assumed to define the movement direction. Here, we show experimentally that the sperm head movement direction is not opposite to the wave propagation direction when sperm move in a viscoelastic fluid next to a solid substrate. The oscillation of the movement direction is out of phase with the oscillation of the wave direction, and the movement has a larger amplitude than the wave direction. When we tried to reconstruct the movement direction as a linear combination of the head orientation and the wave direction, we found that the contributions from these two changed in time. Further, the last bend of the flagellum does not move in the lab frame (as observed under the microscope) on a deformable polydimethylsiloxane (PDMS) substrate, which likely results from the flagellum indenting the surface, providing a new mechanism for locomotion. Overall, our results highlight the appearance of the sperm head in steering motility and provide quantitative tests on sperm chemotaxis and flagellar bending mechanism.

Across length scales, living organisms often need to move in a fluid to perform physiological functions[1]. Due to their sizes, microorganisms, such as bacteria, algae, and sperm, exist under the low Reynolds number limit ($Re \ll 1$), where the fluid flow generated by a boundary movement reaches its laminar steady state almost instantaneously, the reverse motion of the boundary generates the exact opposite of the flow, and the flow speed is exactly proportional to the speed of the boundary movement[2]. Due to this linearity, back-and-forth motion leads to no net displacement of the organism[3].

In terms of the microscopic, mechanistic mechanism, it is understood that microorganisms drive fluid movement through a no-slip boundary on the flagellum, according to resistive force theory[4,5]. The key to this is a propagating wave along the flagellum that breaks the cancellation between back-and-forth motions, leading to the description that the cell is driven toward the opposite direction of the wave[6–9]. In mammalian sperm, a helical wave has been reported in watery lab medium[10], while more planarized beating has been reported in an environment more similar to what they encounter in the female reproductive tract, i.e., at the interface of viscoelastic mucus and solid tissue[11,12]. The propagating wave is generated by active bending of the flagellum throughout the whole length in sperm[13,14]. There are no known motors outside the axoneme, and the sperm head is assumed to be a passive carriage pushed by the flagellum, with the head itself thought to influence motility only in "fine structure"[4].

The comparison between resistive force theory and the experimental measurements has not been perfect, as the direction of the movement can differ by 15-35%[4], in part because sperm mostly move near the interface of the fluid and the solid[15,16]. On the fluid front, slippage is commonly observed near a solid boundary for a polymer solution[17], which is not considered in typical computational models. On the solid front, flocking kind of long-range collective behaviors have been reported[18–20]. At the low Reynolds number limit, flocking was theoretically found to be unstable when the momentum between the self-propelled particles and the surrounding fluid needs to be conserved[21], making the observation an indication of direct momentum transfer between sperm and solid substrate. We have previously visualized the interaction between the flagellum and substrate directly[22]. Also, friction-based locomotion has been analyzed for snakes[23], and the general movement direction does not appear to be much different from the opposite of the overall wave direction.

For microswimmers, the ability to control the randomization of the movement direction is an integral part of their motility mechanism. In *Escherichia coli*, when at least one flagellum rotates in the opposite direction, the whole cell stops and "tumbles", randomly reorienting at a moment[24,25]. In *Vibrio alginolyticus*, a buckling at the base of the flagellum[26] results in a "flick" motion[27], which also randomizes orientation instantaneously. In mammalian sperm, however, no such single-event direction randomization is known. It has been thought that sperm change movement direction by changing the direction of the flagellum[4]. We aim to examine this experimentally.

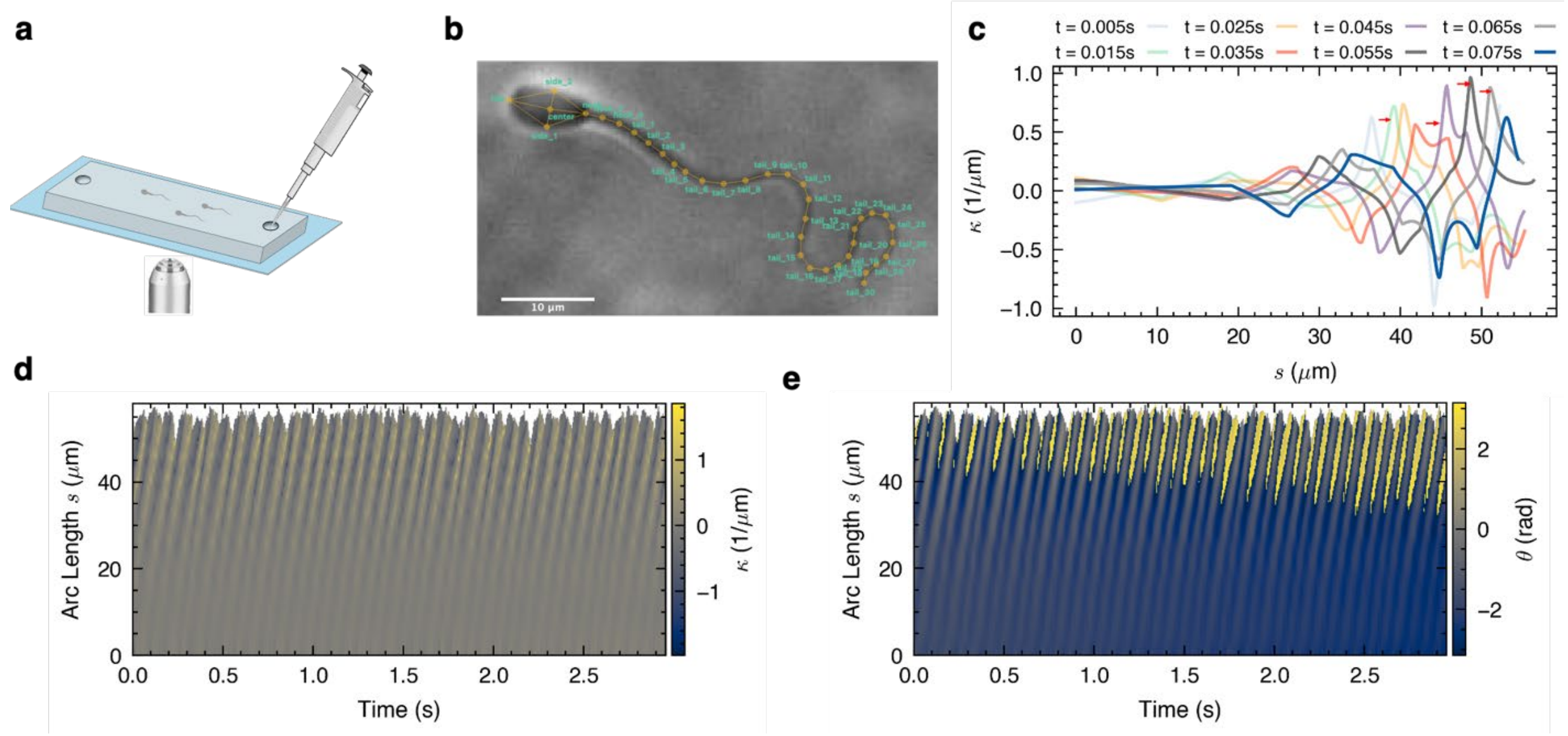

**Fig. 1 | Extraction of the sperm flagellar beating. a** Rendering of experimental setup. Sperm were seeded into a polydimethylsiloxane (PDMS) microfluidic device cast from a silicon wafer patterned with SU-8 and filled with 1% methylcellulose (MC) in Tyrode's Albumin Lactate Pyruvate (TALP). The channel is approximately $100\,\mu$m deep. **b** Skeleton for SLEAP tracking overlaid onto a still image of bovine sperm frame. The skeleton included five points that define the sperm head and 32 points along the beating flagellum. **c** $\kappa$ vs. $s$ throughout a beating cycle. **d** Kymograph of $\kappa(s,t)$. Beyond the midpiece region, the slope of the stripes remains constant, consistent with the description of a $\kappa$ wave propagating along $s$. **e** Kymograph of $\theta(s,t)$. It can again be seen that beyond the midpiece region, the slope of the stripes remains constant, showing a wave propagating along $s$. Due to the turning of the cell, there is a drifting mean.

# Results

## Extraction of the beating of sperm flagellum

Sperm thawed from a frozen bovine sperm sample were seeded into a microfluidic device through a seeding port and imaged on an inverted microscope as shown in Fig. 1a. The channel within the device was filled with 1% methylcellulose dissolved in standard bovine sperm medium. Videos were recorded at 200 frames per second with a 20X objective, with the camera focused on cells moving at the upper surface (PDMS) of the channel. To facilitate the analysis of the movement of the head and flagellum, we employed the machine learning tool SLEAP[28] . A model was manually trained to automatically track cells throughout the videos by placing a skeleton of 5 points representing the head and 32 points representing the flagellum for each time point $t$ [Fig. 1b]. After tracking, exported tail points were used as nodes to fit second-order B-splines with 4000-5000 points[29] , and the resulting curve was resampled into fixed increments of $0.1\,\mu$m in the arc length from the head $s$. The angle spanned between the flagellum and the $x$ axis, $\theta(s,t)$, and the curvature $\kappa(s,t) = \partial\theta/\partial s$, were then computed, as shown in Fig. 1c-d. For comparison to literature[30], we also present $\theta(s,t)$ in Fig. 1e.

## Forward Head Movement of the Sperm and the Opposite of the Wave Directions Differ

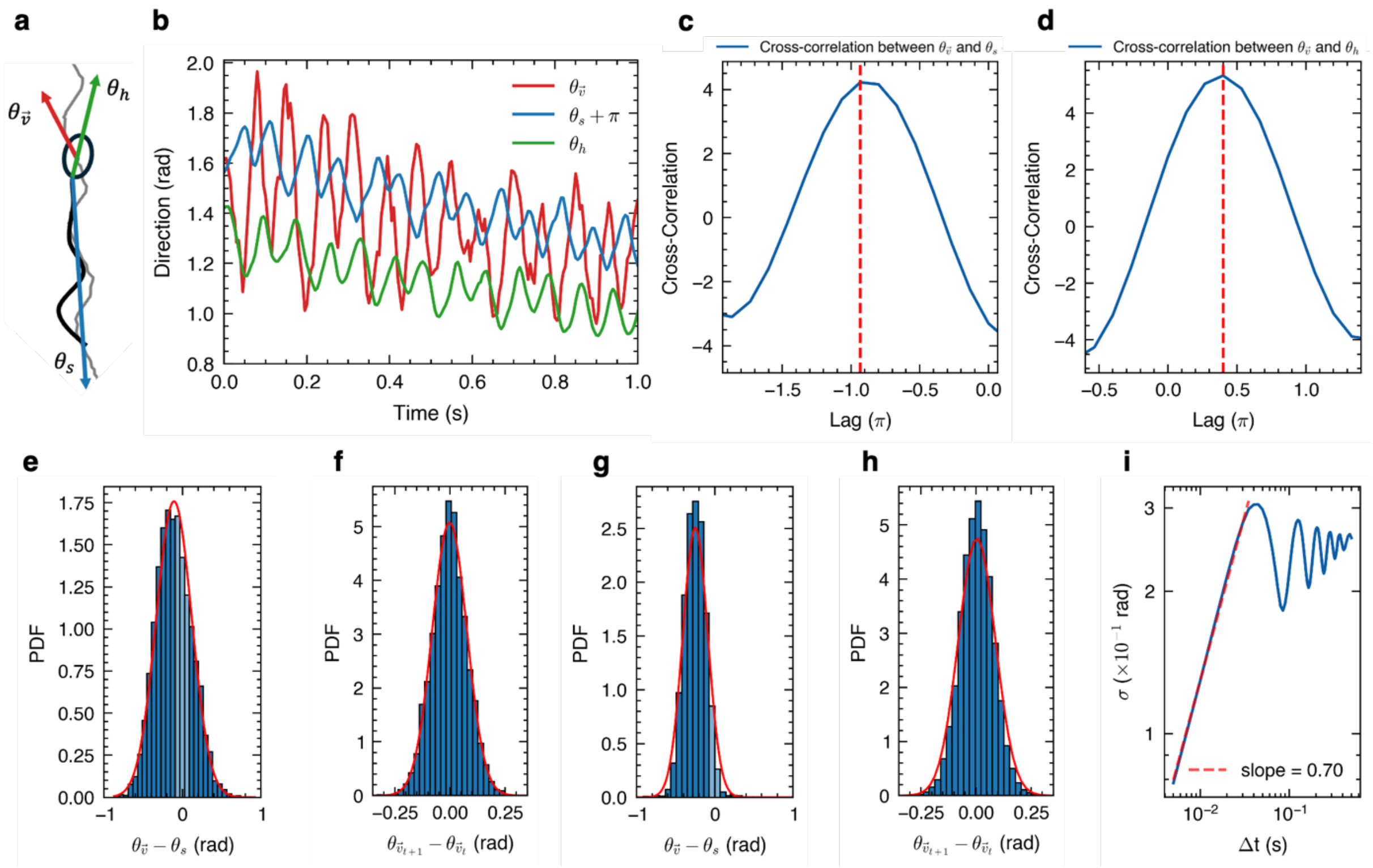

**Fig. 2 | Fundamental discrepancies exist between the movement direction of the sperm head and the opposite of the wave propagation direction. a** Directions of cell movement ($\theta_{\vec{v}}$), head orientation ($\theta_h$), and wave propagation ($\theta_s$) overlaid onto a single-cell trajectory (gray line). **b** Different directions throughout a trajectory. $\pi$ is added to $\theta_s$ to shift it closer to $\theta_{\vec{v}}$ and $\theta_h$ for easier comparison, so that if $\theta_s$ and $\theta_{\vec{v}}$ were exactly opposite, they would have overlapped perfectly. The amplitude difference between $\theta_{\vec{v}}$ and $\theta_s$ illustrates the disparity between the two directions. Also, $\theta_{\vec{v}}$ and $\theta_s$ appear out of phase. **c** Cross correlation function between $\theta_{\vec{v}}$ and $\theta_s$ peaked with a lag of $-(0.93 \pm 0.07)\pi$, indicating the two oscillations are mostly out of phase. **d** Cross correlation function between $\theta_{\vec{v}}$ and $\theta_h$ peaked with a lag of $(0.40 \pm 0.07)\pi$. **e** Aggregate PDF of $\theta_{\vec{v}} - \theta_s$ across 17 trajectories. Counts with $|\theta_{\vec{v}} - \theta_s| \leq 5°$ are shown in lighter blue. The overwhelming majority of counts show a larger difference between the two directions. The distribution is a Gaussian with $\mu = -0.11$ (rad) and $\sigma = 0.23$ (rad). The non-zero mean stems from differences in the y-intercept of the linear drift of the time series. **f** Aggregate PDF of $\theta_{\vec{v}_{i+1}} - \theta_{\vec{v}_i}$. Again, the distribution is a Gaussian except with a $\mu = 0.00$ and $\sigma = 0.08$ (rad). **g** The same analysis as **e** for a long trajectory (>10000 frames). **h** The same analysis as **f** for a long trajectory. **i** $\sigma_{\theta(t+\Delta t)-\theta(t)}$ as a function of $\Delta t$. At short time scale, a power relation $\sigma \sim (\Delta t)^a$, with $a = 0.70$, between diffusive and ballistic, was observed.

To determine the direction of the wave propagation seen in Fig. 1c, we observed that the wave propagated along $s$ at a constant speed $v_{wave}$, as represented by the constant slope of the stripes beyond the midpiece region of the flagellum ($s \geq 15$ $\mu$m) in Fig. 1d. This implies $\kappa(s,t) = \kappa(s - v_{wave}t)$. Similar $\kappa(s,t)$ or $\theta(s,t)$ have been previously reported[30–32]. As the curvature wave propagated down $s$, the direction of the wave was the vector sum of all $d\vec{s}$, which was the direction of $\vec{s}$ between the end of the midpiece and the end of the flagellum. The angle between $\vec{s}$ and the $x$ axis is denoted as $\theta_s$ [Fig. 2a]. A method based on conserving the waveform between $\kappa(s,t)$ and $\kappa(s - v_{wave}\delta t, t + \delta t)$ can be found in Supplementary Materials.

The head movement direction angle, $\theta_{\vec{v}}$, is defined by the vector connecting the consecutive center head points [Fig. 2a and Supplementary Video 1]. The comparison between the directions is shown in Fig. 2b. $\pi$ was added to the wave direction so that if it were exactly opposite to the movement direction, we should have seen two overlapping curves. This was clearly not observed. In many cases, the amplitude for the $\theta_{\vec{v}}$ oscillation was found to be larger than that of $\theta_s$. Since a sperm is made up of a head and a tail, we included the head orientation angle $\theta_h$ for comparison, which was defined by the vector from the point where the tail connects to the head to the top of the head, with an amplitude similar to that of $\theta_s$. Besides the differences in values and oscillation amplitudes, $\theta_s$ appeared roughly out of phase with respect to $\theta_{\vec{v}}$, with the highest cross correlation between the two found with a lag of $-(0.93 \pm 0.07)\pi$ [Fig. 2c], while $\theta_h$ appeared more in phase with $\theta_{\vec{v}}$, with cross correlation peaks at $(0.40 \pm 0.07)\pi$ phase difference [Fig. 2d]. This last observation suggested that the head orientation tracks the movement direction more faithfully than the wave propagation direction on the tail.

We next examined the differences between $\theta_{\vec{v}}$ and $\theta_s$. Across an ensemble of cells, we report the histogram of $\theta_{\vec{v}} - \theta_s$ in Fig. 2e. The vast majority of counts had $\theta_{\vec{v}}$ and $\theta_s$ varying notably. The distribution of the differences closely resembled a Gaussian. The non-zero mean can be attributed to the difference in $y$-intercept between the linear drifts of $\theta_{\vec{v}}$ and $\theta_s$. The non-zero mean disappeared when we removed the constant drift in angles, making the oscillation about zero.

If we examined the change in $\theta_{\vec{v}}$ between consecutive frames, another Gaussian appeared [Fig. 2f]. Similar behaviors were seen when the same analyses were performed on a single cell over a ~50 s (>10000 frames) trajectory [Fig. 2g-h]. The width of the Gaussians in Fig. 2f, h, $\sigma$, increases as the time between frames, $\Delta t$, increases at short time scale, following a power increase [Fig. 2i], before the oscillation became significant at $\Delta t > 0.03$ s. This power relation indicates that (1) the Gaussian reported in Fig. 2e was not an artifact from the analysis; (2) non-negligible diffusive-like randomization occurred at short time scale.

**Dynamic Contribution of Head and Tail to the $\vec{v}$ Direction**

Morphologically, mammalian sperm consist of a head and a flagellum. Therefore, at first order, we examined $\theta_{\vec{v}}$ as a linear combination of $\theta_s$ and $\theta_h$, since (1) the linearity of Stokes flow; (2) once the linear drift was removed, all the time series were approximately sinusoidal oscillations, and one sinusoidal oscillation can be expressed as a linear combination of two others with the same frequency. By fitting

$$\theta_{\vec{v}} = \alpha\theta_s + \beta\theta_h \quad (1)$$

for each beating cycle [mean period obtained from Fourier analysis of $\kappa(s,t)$], we found that the method produced both good and poor fits [Fig. 3a]. Some of the poor fits can be attributed to different frequencies between different oscillations [Fig. 3a, middle], while some have $\theta_s$ and $\theta_h$ in a phase relation that cannot generate the phase of $\theta_{\vec{v}}$ [Fig. 3a, right].

In Fig. 3b, we show that (1) $\alpha$ and $\beta$ varied from cycle to cycle; (2) $\beta$ were more likely to be positive while $\alpha$ were more likely to be negative, which were consistent with our previous findings about the phase relations between $\theta_{\vec{v}}$, $\theta_h$, and $\theta_s$. Fig. 3c further illustrates the sign of $\alpha$ and $\beta$, and that the values of $\alpha$ and $\beta$ were not correlated with the quality of the fitting. $\alpha$ and $\beta$ also were not correlated to each other [Fig. 3d]. If we only take data points with higher $R^2$, the difference in sign tendency would be more distinct. That is, the better the fit, the more obvious it is that the head orientation positively contributed to the head movement direction.

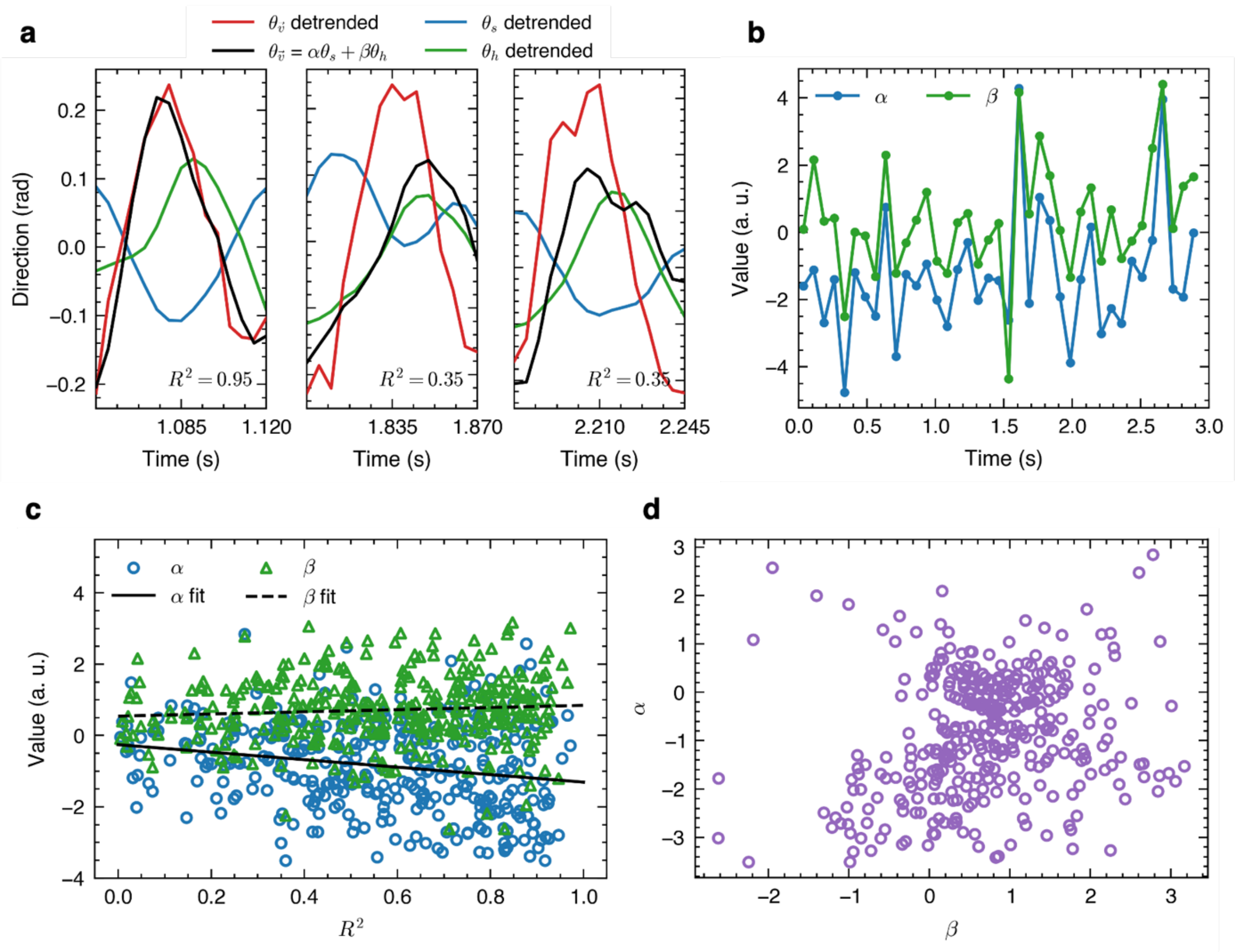

**Fig. 3 | Reconstruction of $\theta_{\vec{v}}$ from the dynamics of $\theta_h$ and $\theta_s$ by fitting Eq. [1]. a** Fitting for $\alpha$ and $\beta$, each beating cycle yielded a reconstructed time series (black line). Some cycles were better fitted than others. **b** $\alpha$ and $\beta$ as a function of time for a single trajectory. Each parameter appeared to vary independently of the other. **c** $\alpha$ and $\beta$ as a function of $R^2$. There was no clear indication that the values of the parameters correlate with the fitting quality. **d** Scatter plot between $\alpha$ and $\beta$. No clear correlation was seen between the two variables.

## Stationary Last Bend of the Flagellum

Here, we report that the last bend of the flagellum (where the largest curvature was found) did not move in the lab frame (as observed under the microscope), while the tail slid along the fixed contour of the last bend until the distal end of the tail slid past the bend [Fig. 4a and Supplementary Video 2]. This appearance of "stalling" was observed on the interface of the polymer solution and PDMS.

We characterized this quantitatively as a kinematic, semi-holonomic constraint on the speeds related to the location of the last bend. Let us focus on the point on the flagellum with the largest curvature around the last bend, as indicated by the red arrows in Fig. 4a. We could not measure the speed the flagellum slid forward along the contour independently, but by the assumption that the fluctuation in length between this point and the distal end of the flagellum was minimal, we took the speed of the distal end ($v_{tail\,end}$) to approximate it.

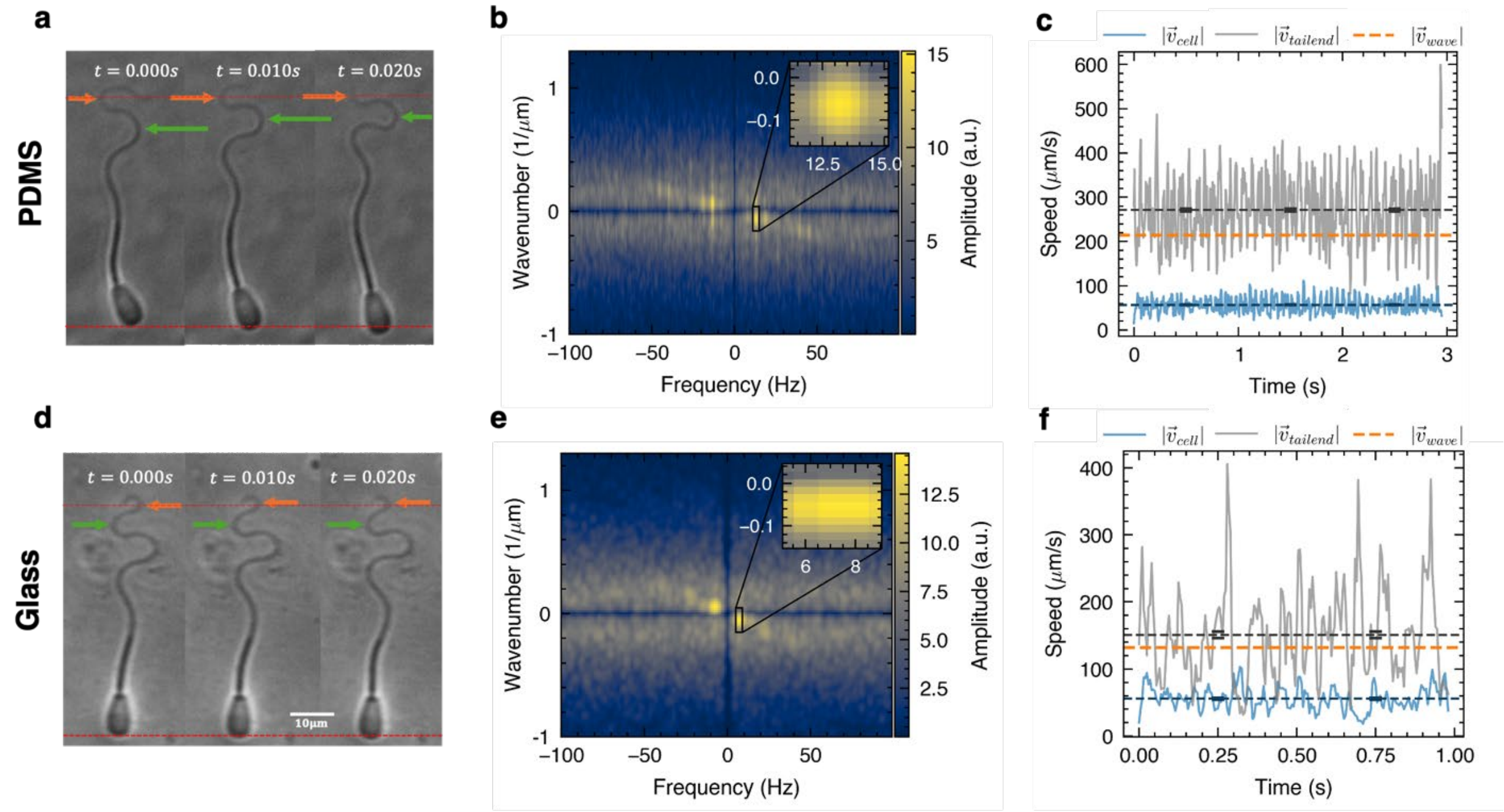

**Fig. 4 | Characterization of the observed lack of movement of the last bend of the flagellum.** **a** Montage of sperm movement close to a PDMS substrate. As the head moves forward and the proximal bend of the flagellum propagates rearward, the distal bend remains stationary in the lab frame. **b** Power spectrum of the flagellar wave at a PDMS substrate in log scale. Taking the two-dimensional Fourier transformation of the $\kappa(s,t)$ kymograph reveals the power spectrum, whose dominant peak location yields $v_{wave}$. **c** Speed of the wave, cell, and last bend as a function of time for a cell close to a PDMS substrate. The speed of the last bend is approximately equal to the sum of $v_{wave}$ and the cell speed, meaning that the constraint described by $v_{last\,bend} \approx v_{\text{wave}} + v_{\text{cell}}$ is a good approximation for the last bend to appear nearly stationary in the lab frame. SEM noted at 0.5, 1.5, and 2.5 seconds. **d** Montage of sperm movement close to a glass substrate. The last bend is observed to move backward slightly. **e** The same analysis as in **b** close to a glass substrate. **f** The same analysis as **c** close to a glass substrate. SEM noted at 0.25 and 0.75 seconds.

The speed that the wave propagates backward along $s$, $v_{wave}$, was extracted through a two-dimensional Fourier analysis on $\kappa(s,t)$, determined from the dominant peak location in the two-dimensional power spectrum with respect to the $k_s$ and $\omega$ axes [Fig. 4b]. At the point of maximum curvature, the wave propagation velocity pointed backward while the sliding velocity pointed forward. The difference between $v_{tail\,end}$ and $v_{wave}$ must be the speed of the sperm (head) moving forward, $v_{cell}$, which we measured independently. Overall, the relationship between the three speeds could be written approximately as

$$v_{tail\,end} - v_{wave} \approx v_{cell} \tag{2}$$

which is illustrated in Fig. 4c.

The stationary flagellum in the lab frame indicated that the resistance the flagellum experienced normal to contour direction was much larger than the resistance tangential to the contour, which could not solely come from the fluid, with a ratio at 1.6-2.0[4] . We suspected that part of the force normal to the flagellum was a result of the flagellum pushing into the deformable PDMS substrate (Young's modulus of ~ 2 MPa[33,34]), or direct indentation by the flagellum. To test this, we performed the same analyses on a glass surface, which presumably cannot be deformed. In this case, we found the last bend moving backward, and the constraint Eq. [2] was no longer satisfied, supporting that indentation of the surface is the source of the stationary last bend of the flagellum.

# Discussion

Here, we report that the sperm head movement direction is different from the opposite of the wave direction, as seen by the amplitude of the movement oscillation being larger than the tail wave, and the two oscillations being almost out of phase. The simple model relying solely on a propagating wave is not supported by the experimental data to actually describe sperm motility.

The origin of the discrepancies is curious. In the condition of this study, when sperm moved along a flat upper PDMS surface, we previously showed that there was a slithering kind of motion by the sperm flagellum[22]. In the study of snake slithering, it has been shown that periodic redistribution of the weight of the body increases the oscillation amplitude (although not larger than the experimental measurements) than an evenly distributed weight[23]. Clearly, to explain the discrepancies we observed, a mechanism other than the third-dimensional movement of the flagellum (perpendicular to the solid surface) is needed.

Here, we present data indicating that sperm head orientation appears to approximately steer the movement of the cell. This is conceptually counterintuitive in two ways. First, this description is against the fluid mechanics descriptions seeing the head as a passive object pushed by the flagellum. Second, there is no known motor protein or other active element in directly moving the sperm head. On the hydrodynamics front, we note that it is generally expected for the settling elongated object to have the long axis parallel to the settling direction[35–37], which supports our observation that the elongated sperm head can steer sperm movement through hydrodynamics, which is also in agreement with the resistive force theory[4]. This also provides a basis for different species to have different head shapes through fine-tuning steering parameters. On the motor front, the sperm head clearly cannot take up an active steering role by itself, but it has been reported that the base where the flagellum and head connect can undergo kinking motion to allow the head a different orientation[38]. Overall, although the sperm head does not actively perform steering, our data show that the activities of the axoneme somehow effectively make the head orientation a determinant in sperm movement direction.

In single-flagellum bacteria with a cell body acting like the sperm head, it has been recognized that significant reorientation happens when the hook that connects the flagellum to the body buckles[26]. When they move forward or backward, the cell body and the tail are mostly aligned. When the buckling occurs, the body acquires a new direction, and then the tail moves into the direction set forth by the body[27,39]. Here, we report that the oscillation of the movement direction is mostly in phase with the oscillation of the head orientation, and the head orientations mostly have positive contributions to the movement directions. In both cases, when the sperm head (or the body of the bacteria) and the (opposite of the) tail directions diverge, the movement is seen as driven by the head (or the cell body). In both cases of bacteria and sperm, there is no motor actively turning the head (or the cell body).

Another significance of this comparison is that buckling in single-flagellum bacteria provides a reorientation mechanism, and this reorientation is seen as a crucial component of their navigation mechanism, such as chemotaxis. Although sperm do not exhibit "flick"-style reorientation events, Fig. 2e indicates that the deviation between the movement and head/tail directions or the changes in sequential movement directions in sperm also serves as a reorientation mechanism. The distribution of the change in angle of the movement direction has a width of 0.08 (rad), compared to $4 \times 10^{-9}$ (rad) from rotational diffusion, using the diffusion coefficient formula from Ref. 40 (Diffusion coefficient from Ref. 41,42 yield smaller values). The superdiffusion observed in Fig. 2i also indicates central limit theorem style of randomization in short time scales. Here, we provide several measures for future studies in an environment with a non-neutral environment, such as under a chemoattractant gradient. First, if the frame-to-frame movement direction change does reflect the reorientation, we expect to see the peak to narrow under such a gradient. Second, since those are modulated by the head and the tail, we expect the dynamic modulation of $\alpha$ and $\beta$ to have more correlated features.

We also showed that the movement direction cannot be fully recovered from the linear combination of the dynamics of the head and wave alone. The majority of the bad fits show different frequencies among the head/wave/movement. How much of this can be explained by the 3D beating remains to be examined.

Lastly, we report the observation that the last bend of the flagellum does not move in the lab frame on a PDMS surface and show that the same is not found on a glass surface. The comparison between PDMS and glass supports the idea that the PDMS surface is indented by the flagellum, which provides normal force preventing the wave from moving backward from the head. Note that the diameter of the flagellum tapers down to 200 nm toward the distal end[43], making the depth of indentation needed for producing significant normal force to the flagellum much less. Given that the wave propagates backward on the glass surface, the presumed normal force from indentation around the straight portion of the flagellum after the last bend may be another source of locomotion, which would be in agreement with a report in human sperm[44]. Since sperm flagellum is known to be locally bent by the dynein motors[13,14], we assume that the reported kinematic, semi-holonomic constraint is a result of the self-organization between the dynein motors and the reaction-diffusion regulation in the flagellum[45]. This potentially provides us with an additional venue for testing models for the relationship between molecular motor coordination and flagellar bending[46,47]. The indentation also further highlights the significance and complexity of the solid-solid interactions between the sperm and the solid substrate[22]. As solid-solid interaction has been known as a source of entropy generation and break the time-reversibility of Stokes flow[48], this interaction also provides a ready answer to how sperm can move in a direction different from the opposite of the wave propagation direction on the flagellum. Improved 3D imaging will help us better model the solid-solid interaction in the future.

# Author Information

**Contributions**

Conceptualization, T.A.-R., K.I., and C.-K.T.; Methodology, R.L.S., L.A. and C.K.T.; Software and Formal Analysis, R.L.S., and C.-K.T.; Investigation, R.L.S., S.U., A.B., and K.I.; Writing-Original Draft, R.L.S.; Review, & Editing: R.L.S., S.U., A.B., L.A., T.A.-R., K.I., C.-K.T.; Supervision, C.-K.T.; Funding Acquisition, C.-K.T.

# Acknowledgments

CKT thanks Chungwei Lin, Donald L. Koch, and Wayne Zhao for useful discussions. This work was supported by NSF DMR 2144064 to CKT. Frozen bovine sperm samples were kindly provided by Select Sires, Inc. The fabrication was performed using the NCSU Nanofabrication Facility (NNF), a member of the North Carolina Research Triangle Nanotechnology Network (RTNN), which is supported by the NSF Grant ECCS-1542015 as part of the National Nanotechnology Coordinated Infrastructure (NNCI).

# Methods

## Microfluidic Device Construction

The design of the silicon mold used in this project is the same as in previous works[22,49]. The channels were 4 cm long, 2.47 mm wide, and 100 $\mu$m deep. The mold was fabricated with a single layer of SU-8 negative photoresist using standard photolithography. SU-8 was selected over etching to ensure clean, sharp channel edges. All fabrication steps were completed using the instruments and cleanroom facilities at North Carolina State University (NNF). Microfluidic devices were cast using PDMS at a 10:1 ratio of base to curing agent; for each device, 13 g of base and 1.3 g of curing agent were combined. The base and curing agent were mixed thoroughly and degassed until all air bubbles were removed. The mixture was poured onto the mold, degassed a second time, and cured at 65 °C for 2 hours. After curing, devices were cut from the mold, and sperm-seeding and fluid-input ports were created using biopsy punches. The PDMS components were bonded to glass slides following oxygen plasma treatment at high RF power for 60 seconds. After bonding, the channel was flushed with ethanol to sterilize the interior and then rinsed thoroughly with deionized water. For experiments, channels were filled with media equilibrated at 38.5 °C under 5% $CO_2$ in humidified air for at least 2 hours, then returned to the incubator until ready for microscopy. During imaging, devices were placed in an environmentally controlled chamber maintained at 38.5 °C. Sperm were seeded into one end of each device immediately before recording.

## Media Preparation

The viscoelastic fluid used in this experiment was 1% MC dissolved in Tyrode's Albumin Lactate Pyruvate (TALP) by weight. Methylcellulose was added to increase viscosity and elasticity so that the fluid more closely simulated conditions within the female reproductive tract [50]. TALP consisted of 99 mM NaCl, 3.1 mM KCl, 0.39 mM $NaH_2PO_4$, 25 mM $NaHCO_3$, 10 mM HEPES free acid, 2 mM $CaCl_2$, 1.1 mM $MgCl_2$, 25.4 mM sodium lactate, 1 mM/mL sodium pyruvate, 5 mg/mL gentamicin, and 6 mg/mL bovine serum albumin (BSA). The solution was titrated to pH 7.4 using 1 M NaOH.

## Sperm Sample Preparation

Frozen bovine sperm samples stored in 500 $\mu$L plastic straws were donated by Select Sires and kept in liquid nitrogen until use. Straws were thawed by placing them in a 37 °C water bath for 30 seconds. The sample was centrifuged through two layers (40% and 80%) of Bovipure in Bovi-Dilute solution at 300 × g for 10 minutes. Solutions were equilibrated at 38.5 °C under 5% $CO_2$ in humidified air for at least 2 hours before centrifugation. Following centrifugation, the supernatant was removed and the sperm pellet was resuspended in 3 mL TALP. The suspension was centrifuged again at 300 × g for 3 minutes. The supernatant was removed once more and the pellet was resuspended in 300 $\mu$L TALP, then placed in an incubator at 38.5 °C under 5% $CO_2$ in humidified air until use.

**Recording**

Imaging was performed using a Nikon Eclipse inverted phase contrast microscope with a 20× objective and a Hamamatsu ORCA Flash 4.0 V3 camera. Approximately 10 $\mu$L of diluted sperm solution was introduced into a microfluidic device filled with 1% MC and placed in a temperature-controlled chamber on the microscope. After allowing time for sperm to migrate away from the entry port, recordings were acquired using NIS Elements BR software. The camera was focused on the top surface of the microfluidic device to maintain a clean background and minimize optical obstruction. A wide field of view was used to keep cells in frame for long trajectories. Each video was recorded at approximately 200 frames per second.

**Machine-Learning-Based Tracking**

Sperm cells were tracked automatically using SLEAP, a machine learning framework for multi-animal pose estimation [28]. Recorded videos were converted from .nd2 to .avi format before being imported into SLEAP. A custom skeleton was created to mark anatomical features of interest, including one node at the center of the sperm head, nodes on the top, left, and right sides of the head, three nodes along the midpiece region, and 30 nodes distributed along the rest of the flagellum [Fig. 1b]. A training dataset was generated by selecting frames at random and manually adjusting the skeleton in each frame to match the cell's position and orientation.

The model was trained using the multi-animal top-down approach, in which the software first identified a centroid for each cell, cropped around it, and then predicted node positions using a centered-instance confidence map. The centroid was defined as the node at the center of the head. Model performance was refined through repeated cycles of training and correction. To optimize accuracy, the crop size and input scaling were tuned so that the entire cell remained visible while minimizing unnecessary background. Rotation correction was enabled to allow tracking of cells moving in any direction. For the centered-instance model, the batch size and plateau patience were increased to extend training. Hard keypoints were assigned to nodes near the distal end of the flagellum, where the model initially performed poorly, so that greater weight was placed on improving accuracy in that region. After refinement, the model accurately placed nodes along the entire cell. An inference step was run to track all cells and assign each a unique track identifier. After manually confirming node placement accuracy, model output for each video was exported as a .csv file, and a separate .txt file containing node positions was generated for each tracked cell in each frame.

**Flagellar Analysis**

After exporting tracking data from SLEAP, custom Python scripts were used to extract and analyze trajectories. The first script read each .csv file, removed accuracy scores and unused metadata, and organized the remaining coordinates into continuous trajectories for individual cells. Tracks shorter than 300 frames were discarded. For each cell and frame, a .txt file containing the positions of all nodes was generated. The second script reconstructed the flagellum waveform by fitting a smooth B-spline to the $(x, y)$ node positions for each frame, producing a continuous curve parameterized by $l \in [0,1]$. The spline was evaluated at 1000 closely spaced values of $l$ to produce a dense curve from which the cumulative arc length $s$ was computed, giving the true physical distance along the flagellum. Interpolation was then used to map any value of $s$ to the corresponding point on the spline, and the curve was resampled at increments of $0.1$ $\mu$m to produce evenly spaced $(x, y)$ points along the flagellum. The third script performed all flagellar analyses, loading the resampled coordinate files, computing the desired metrics, and saving results to text files.

# References

1. Gazzola, M., Argentina, M. & Mahadevan, L. Scaling macroscopic aquatic locomotion. *Nat. Phys.* **10**, 758–761 (2014).
2. Lauga, E. Life around the scallop theorem. *Soft Matter* **7**, (2010).
3. Purcell, E. M. Life at low Reynolds number. *Am. J. Phys.* **45**, 3–11 (1977).
4. Friedrich, B. M., Riedel-Kruse, I. H., Howard, J. & Jülicher, F. High-precision tracking of sperm swimming fine structure provides strong test of resistive force theory. *J. Exp. Biol.* **213**, 1226–1234 (2010).
5. Keaveny, E. E. & Brown, A. E. X. Predicting path from undulations for C. elegans using linear and nonlinear resistive force theory. *Phys. Biol.* **14**, 025001 (2017).
6. Elgeti, J., Winkler, R. & Gompper, G. Physics of Microswimmers - Single Particle Motion and Collective Behavior. *Rep. Prog. Phys.* **78**, 056601 (2015).
7. Liu, B., Chen, L. & Xu, W. Effects of flagellar morphology on swimming performance and directional control in microswimmers. *Phys. Fluids* **37**, 041912 (2025).
8. Puri, D. B., Wadhwa, H., Hemadri, V., Biswas, G. & Tripathi, S. Rheotaxis of sperm cells: mechanism and applications in microfluidic separation. *Discov. Fluid Mech.* **2**, 1 (2026).
9. Cheng, S. *et al.* Corkscrew motion of Trypanosoma brucei is driven by helical beating of the flagellum and facilitated by its bent shape. *Proc. Natl. Acad. Sci.* **123**, e2536746123 (2026).

10. Corkidi, G. *et al.* Human sperm rotate with a conserved direction during free swimming in four dimensions. *J. Cell Sci.* **136**, jcs261306 (2023).

11. Nosrati, R., Driouchi, A., Yip, C. M. & Sinton, D. Two-dimensional slither swimming of sperm within a micrometre of a surface. *Nat. Commun.* **6**, 8703 (2015).

12. Tung, C. *et al.* Fluid viscoelasticity promotes collective swimming of sperm. *Sci. Rep.* **7**, 3152 (2017).

13. Machin, K. E. Wave Propagation Along Flagella. *J. Exp. Biol.* **35**, 796–806 (1958).

14. Lin, J. & Nicastro, D. Asymmetric distribution and spatial switching of dynein activity generates ciliary motility. *Science* **360**, (2018).

15. Rothchild. Non-random Distribution of Bull Spermatozoa in a Drop of Sperm Suspension. *Nature* **198**, 1221–1222 (1963).

16. Denissenko, P., Kantsler, V., Smith, D. J. & Kirkman-Brown, J. Human spermatozoa migration in microchannels reveals boundary-following navigation. *Proc. Natl. Acad. Sci.* **109**, 8007–8010 (2012).

17. Barnes, H. A. A review of the slip (wall depletion) of polymer solutions, emulsions and particle suspensions in viscometers: its cause, character, and cure. *J. Non-Newton. Fluid Mech.* **56**, 221–251 (1995).

18. Rothschild, Lord. The Activity of Ram Spermatozoa. *J. Exp. Biol.* **25**, 219–226 (1948).

19. Creppy, A. *et al.* Symmetry-breaking phase transitions in highly concentrated semen. *J. R. Soc. Interface* **13**, 20160575 (2016).

20. Packard, C. R. *et al.* Self-organized vortex phases and hydrodynamic interactions in Bos taurus sperm cells. *Phys. Rev. E* **110**, 014407 (2024).

21. Aditi Simha, R. & Ramaswamy, S. Hydrodynamic Fluctuations and Instabilities in Ordered Suspensions of Self-Propelled Particles. *Phys. Rev. Lett.* **89**, 058101 (2002).

22. Unnikrishnan, S. *et al.* Hybrid motility mechanism of sperm at viscoelastic fluid-solid interface. *Sci. Rep.* **14**, 21841 (2024).

23. Hu, D. L., Nirody, J., Scott, T. & Shelley, M. J. The mechanics of slithering locomotion. *Proc. Natl. Acad. Sci.* **106**, 10081–10085 (2009).

24. BERG, H. C. & BROWN, D. A. Chemotaxis in Escherichia coli analysed by Three-dimensional Tracking. *Nature* **239**, 500–504 (1972).

25. Macnab, R. M. & Ornston, M. K. Normal-to-curly flagellar transitions and their role in bacterial tumbling. Stabilization of an alternative quaternary structure by mechanical force. *J. Mol. Biol.* **112**, 1–30 (1977).

26. Son, K., Guasto, J. S. & Stocker, R. Bacteria can exploit a flagellar buckling instability to change direction. *Nat. Phys.* **9**, 494–498 (2013).

27. Xie, L., Altindal, T., Chattopadhyay, S. & Wu, X.-L. Bacterial flagellum as a propeller and as a rudder for efficient chemotaxis. *Proc. Natl. Acad. Sci.* **108**, 2246–2251 (2011).

28. Pereira, T. D. *et al.* SLEAP: A deep learning system for multi-animal pose tracking. *Nat. Methods* **19**, 486–495 (2022).

29. Mary, H. & Brouhard, G. J. Kappa ($\kappa$): Analysis of Curvature in Biological Image Data using B-splines. *bioRxiv* 852772 (2019) doi:10.1101/852772.

30. Ma, R., Klindt, G. S., Riedel-Kruse, I. H., Jülicher, F. & Friedrich, B. M. Active Phase and Amplitude Fluctuations of Flagellar Beating. *Phys Rev Lett* **113**, 048101 (2014).

31. Gaffney, E. A., Gadêlha, H., Smith, D. J., Blake, J. R. & Kirkman-Brown, J. C. Mammalian Sperm Motility: Observation and Theory. *Annual Review of Fluid Mechanics* vol. 43 501–528 (2011).

32. Sharma, A., Friedrich, B. M. & Geyer, V. F. Active fluctuations of axoneme oscillations scale with number of dynein motors. *Proc. Natl. Acad. Sci.* **121**, e2406244121 (2024).

33. Sales, F. C. P., Ariati, R. M., Noronha, V. T. & Ribeiro, J. E. Mechanical Characterization of PDMS with Different Mixing Ratios. *Procedia Struct. Integr.* **37**, 383–388 (2022).

34. Cho, H. S., Moon, H.-Y., Lee, H. S., Kim, Y. T. & Jeoung, S. C. Formulation Prediction for Young's Modulus of Poly(dimethylsiloxane) by Spectroscopic Methods. *Bull. Korean Chem. Soc.* **42**, 1225–1231 (2021).

35. Leal, L. G. The slow motion of slender rod-like particles in a second-order fluid. *J. Fluid Mech.* **69**, 305–337 (1975).

36. Joseph, D. D. & Liu, Y. J. Orientation of long bodies falling in a viscoelastic liquid. *J. Rheol.* **37**, 961–983 (1993).

37. Dabade, V., Marath, N. K. & Subramanian, G. Effects of inertia and viscoelasticity on sedimenting anisotropic particles. *J. Fluid Mech.* **778**, 133–188 (2015).

38. Khanal, S. *et al.* A dynamic basal complex modulates mammalian sperm movement. *Nat. Commun.* **12**, 3808 (2021).

39. Park, Y., Kim, Y. & Lim, S. Locomotion of a single-flagellated bacterium. *J. Fluid Mech.* **859**, 586–612 (2019).

40. Eisner, A. D. & Gallily, I. On the stochastic nature of the motion of nonspherical aerosol particles: III. the rotational diffusion diadic and applications. *J. Colloid Interface Sci.* **81**, 214–233 (1981).

41. Berg, H. C. *Random Walks in Biology*. (Princeton University Press, 1993). doi:10.2307/j.ctv7r40w6.

42. Li, M., Mulholland, G. W. & Zachariah, M. R. Rotational Diffusion Coefficient (or Rotational Mobility) of a Nanorod in the Free-Molecular Regime. *Aerosol Sci. Technol.* **48**, 139–141 (2014).

43. Lindemann, C. B. & Lesich, K. A. Functional anatomy of the mammalian sperm flagellum. *Cytoskeleton* **73**, 652–669 (2016).

44. Neal, C. V., Hall-McNair, A. L., Kirkman-Brown, J., Smith, D. J. & Gallagher, M. T. Doing more with less: The flagellar end piece enhances the propulsive effectiveness of human spermatozoa. *Phys. Rev. Fluids* **5**, 073101 (2020).

45. Cass, J. F. & Bloomfield-Gadêlha, H. The reaction-diffusion basis of animated patterns in eukaryotic flagella. *Nat. Commun.* **14**, 5638 (2023).

46. Gallagher, M. T., Kirkman-Brown, J. C. & Smith, D. J. Axonemal regulation by curvature explains sperm flagellar waveform modulation. *PNAS Nexus* **2**, pgad072 (2023).

47. Maggi, C. *et al.* Thermodynamic Limits of Sperm Swimming Precision. *PRX Life* **1**, 013003 (2023).

48. Ro, S. *et al.* Model-Free Measurement of Local Entropy Production and Extractable Work in Active Matter. *Phys Rev Lett* **129**, 220601 (2022).

49. Phuyal, S., Suarez, S. S. & Tung, C.-K. Biological benefits of collective swimming of sperm in a viscoelastic fluid. *Front. Cell Dev. Biol.* **Volume 10-2022**, (2022).

50. Achinger, L. *et al.* The Known and Unknown About Female Reproductive Tract Mucus Rheological Properties. *BioEssays* **47**, e70002 (2025).